\documentclass[useAMS]{mn2e}
\usepackage{graphicx}

\def\arcsec{\tt ''}
\def\simgt{\ {\raise-.5ex\hbox{$\buildrel>\over\sim$}}\ }

\begin{document}

\title[WET observations of KUV\,05134+2605 and
PG\,1654+160]{\vspace{-11mm}Amplitude and frequency variability of the
pulsating DB white dwarf stars KUV\,05134+2605 and PG\,1654+160 observed
with the Whole Earth Telescope\vspace{-7mm}}
\author[G. Handler et al.]
       {G.\,Handler$^{1,2}$, D.\,O'Donoghue$^1$, 
M.\,M\"uller$^3$, J.-E.\,Solheim$^4$, J.\,M.\,Gonzalez-Perez$^4$,\cr
F.\,Johannessen$^4$, M.\,Paparo$^5$, B.\,Szeidl$^5$, G.\,Viraghalmy$^5$, 
R.\,Silvotti$^6$, G.\,Vauclair$^7$, \cr N.\,Dolez$^7$, E.\,Pallier$^7$, 
M.\,Chevreton$^8$ D.\,W.\,Kurtz$^{7,9}$, G.\,E.\,Bromage$^9$, M.\,S.\,Cunha$^{10,11}$,
\cr R.\,{\O}stensen$^{12}$, L.\,Fraga$^{13}$, A.\,Kanaan$^{13}$, A.\,Amorim$^{13}$,
O.\,Giovannini$^{14}$, \cr S.\,O.\,Kepler$^{15}$, 
A.\,F.\,M.\,da Costa$^{15}$, R.\,F.\,Anderson$^{16}$, M.\,A.\,Wood$^{17}$,
N.\,Silvestri$^{17}$,\cr E.\,W.\,Klumpe$^{18}$, R.\,F.\,Carlton$^{18}$, 
R.\,H.\,Miller$^{19}$, J.\,P.\,McFarland$^{19}$, A.\,D.\,Grauer$^{20}$, \cr
S.\,D.\,Kawaler$^{21}$, R.\,L.\,Riddle$^{21}$, M.\,D.\,Reed$^{22}$, 
R.\,E.\,Nather$^{23}$, D.\,E.\,Winget$^{23}$, \cr J.\,A.\,Hill$^{23}$,
T.\,S.\,Metcalfe $^{23,24}$, A.\,S.\,Mukadam$^{23}$,
M.\,Kilic$^{23}$, T.\,K.\,Watson$^{25}$, \cr S.\,J.\,Kleinman$^{26}$, 
A.\,Nitta$^{26}$, J.\,A.\,Guzik$^{27}$, P.\,A.\,Bradley$^{27}$,
K.\,Sekiguchi$^{28}$, \cr D.\,J.\,Sullivan$^{29}$, T.\,Sullivan$^{29}$,
R.\,R.\,Shobbrook$^{30, 31}$, X.\,Jiang$^{32}$, P.\,V.\,Birch$^{33}$, \cr
B.\,N.\,Ashoka$^{34}$, S.\,Seetha$^{34}$, 
V.\,Girish$^{34}$, S.\,Joshi$^{35}$, T.\,N.\,Dorokhova$^{36}$, \cr
N.\,I.\,Dorokhov$^{36}$, M.\,C.\,Akan$^{37}$, E.\,G.\,Mei\v{s}tas$^{38}$,
R.\,Janulis$^{38}$, R.\,Kalytis$^{39}$, \cr D.\,Ali\v{s}auskas$^{39}$, 
S.\,K.\,Anguma$^{40}$, P.\,C.\,Kalebwe$^{41}$,
P.\,Moskalik$^{42}$, W.\,Ogloza $^{42,43}$, \cr G.\,Stachowski$^{42}$,
G.\,Pajdosz$^{43}$, S.\,Zola$^{43,44}$
\vspace{-6mm}
\and \\
$^1$ South African Astronomical Observatory, P.O. Box 9, Observatory 7935,
South Africa\\
$^2$ Present address: Institut f\"ur Astronomie,  Universit\"at Wien,
T\"urkenschanzstra\ss e 17, A-1180 Wien, Austria\\
$^3$ Department of Astronomy, University of Cape Town, Rondebosch 7700,
South Africa\\
$^4$ Institutt for Matematiske Realfag, Nordlysobservatoriet, Universitet i
Troms\o, 9000 Troms\o, Norway\\
$^5$ Konkoly Observatory, Box 67, H-1525 Budapest XII, Hungary\\
$^6$ Osservatorio Astronomico di Capodimonte, via Moiariello 16, I-80131 Napoli,
Italy\\
$^7$ Observatoire Midi-Pyr\'{e}n\'{e}es, CNRS/UMR5572, 14 av. E. Belin, 
31400 Toulouse, France\\
$^8$ Observatoire de Paris-Meudon, LESIA, 92195 Meudon, France\\
$^9$ Centre for Astrophysics, University of Central Lancashire, Preston 
PR1 2HE, UK\\
$^{10}$ Centro de Astrof\'\i sica da Universidade do Porto, Rua das
Estrelas, 4150-762 Porto, Portugal\\
$^{11}$ Instituto Superior da Maia, Lugar de Vilarinho, 4470 Castelo da Maia,
Portugal\\
$^{12}$ Isaac Newton Group of Telescopes, 37800 Santa Cruz de La Palma, 
Canary Islands, Spain\\
$^{13}$ Departamento de F\'{\i}sica, Universidade Federal de Santa Catarina, 
CP 476 - CEP 88040-900 Florian\'opolis, SC - Brazil\\
$^{14}$ Departamento de F\'{\i}sica e Qu\'{\i}mica, Universidade de
Caxias do Sul, 95001-970 Caxias do Sul, RS - Brazil\\
$^{15}$ Instituto de F\'{\i}sica, UFRGS, Campus do Vale, C.P. 15051,
Porto Alegre, RS, Brazil\\
$^{16}$ Department of Physics and Astronomy, University of North Carolina,
Chapel Hill, NC 27599-3255, USA\\
$^{17}$ Department of Physics \& Space Sciences and SARA Observatory,
Florida Institute of Technology, Melbourne, FL 32901, USA\\
$^{18}$ Department of Physics and Astronomy, Middle Tennessee State 
University, Murfreesboro, TN 37132, USA\\
$^{19}$ Department of Physics and Astronomy, Georgia State University, 
Atlanta, GA 30303, USA\\
$^{20}$ Department of Physics and Astronomy, University of Arkansas at 
Little Rock, Little Rock, AR 72204, USA\\
$^{21}$ Department of Physics and Astronomy, Iowa State University,
Ames, IA 50011, USA\\
$^{22}$ Department of Physics, Astronomy and Material Science, SW Missouri 
State University, 901 S. National, Springfield, MO 65804 USA\\
$^{23}$ Department of Astronomy and McDonald Observatory, University of Texas
at Austin, Austin, TX 78712, USA\\
$^{24}$ Harvard-Smithsonian Center for Astrophysics, 60 Garden Street,
Cambridge, MA 02138, USA\\
$^{25}$ Information Technology Services Department, Southwestern University,
Georgetown, TX 78626, USA\\
$^{26}$ Sloan Digital Sky Survey, Apache Pt. Observatory, P. O. Box 59,
Sunspot, NM 88349, USA\\
$^{27}$ Los Alamos National Laboratory, X-2, MS T-085, Los Alamos, NM
87545-2345, USA\\
$^{28}$ Subaru Telescope, National Astronomical Observatory of Japan, 
650 North A`oh\={o}k\={u} Place, Hilo, HI 96720, USA\\
$^{29}$ School of Chemical and Physical Sciences, Victoria University
of Wellington, PO Box 600, Wellington, New Zealand\\
$^{30}$ P. O. Box 518, Coonabarabran, N.S.W 2357, Australia\\
$^{31}$ Research School of Astronomy and Astrophysics,
Australian National University, Weston Creek P.O., ACT 2611, Australia\\
$^{32}$ National Astronomical Observatories and Joint Laboratory of Optical 
Astronomy, Chinese Academy of Sciences, Beijing, 100012, China\\
$^{33}$ Perth Observatory, Walnut Rd., Bickley, Western Australia 6076,
Australia\\
$^{34}$ Indian Space Research Organization, Vimanapura PO, Bangalore 560 017, 
India\\
$^{35}$ State Observatory, Manora Peak, Naini Tal 263 129, India\\
$^{36}$ Astronomical Observatory, Odessa State University, Shevchenko Park, 
Odessa 270014, Ukraine\\
$^{37}$ Ege University, Science Faculty, Dept. of Astronomy and Space Sciences,
Bornova 35100 Izmir, Turkey\\
$^{38}$ Institute of Theoretical Physics and Astronomy, Go\v{s}tauto 12, 
Vilnius 2600, Lithuania\\
$^{39}$ Astronomical Observatory of Vilnius University, \v{C}iurlionio 29, 
Vilnius 2009, Lithuania\\
$^{40}$ Department of Physics, Mbarara University of Science and
Technology, P. O. Box 1410, Mbarara, Uganda\\
$^{41}$ Physics Department, The University of Zambia, P. O. Box
32379, Lusaka, Zambia\\
$^{42}$ Copernicus Astronomical Center, ul. Bartycka 18, 00-716 Warsaw, Poland\\
$^{43}$ Mount Suhora Observatory, Krakow Pedagogical University, ul. 
Podchor\v a\. zych 2, 30-084 Cracow, Poland\\
$^{44}$ Astronomical Observatory, Jagiellonian University, ul. Orla 171, 
30-244 Cracow, Poland}
\date{Accepted 2002 nnnn nn.
      Received 2002 nnnn nn;
      in original form 2002 nnnn nn}
\maketitle
\begin{abstract}

We have acquired new time series photometry of the two pulsating DB white 
dwarf stars KUV\,05134+2605 and PG\,1654+160 with the Whole Earth Telescope. 
Additional single-site photometry is also presented. We use all these data 
plus all available archival measurements to study the temporal behaviour 
of the pulsational amplitudes and frequencies of these stars for the first 
time.

We demonstrate that both KUV\,05134+2605 and PG\,1654+160 pulsate in many
modes whose amplitudes are variable in time; some frequency variability of
PG\,1654+160 is also indicated. Beating of multiple pulsation modes cannot
explain our observations; the amplitude variability must therefore be
intrinsic. We cannot find stable modes to be used for determinations of
the evolutionary period changes of the stars. Some of the modes of
PG\,1654+160 appear at the same periods whenever detected. The mean
spacing of these periods ($\approx 40$\,s) suggests that they are probably
caused by non-radial gravity-mode pulsations of spherical degree $\ell=1$.  
If so, PG\,1654+160 has a mass around 0.6 $M_{\sun}$.

The time scales of the amplitude variability of both stars (down to two
weeks) are consistent with theoretical predictions of resonant mode
coupling, a conclusion which might however be affected by the temporal
distribution of our data.

\end{abstract}

\begin{keywords}
stars: variables: other -- stars: variables: ZZ Ceti -- stars:
oscillations -- stars: individual: KUV\,05134+2605  -- stars: individual:
PG\,1654+160
\end{keywords}

{\section{Introduction}}

In recent times it has become clear that amplitude and frequency
variations are common amongst pulsating stars. Various mechanisms for
their explanation have been proposed. For example, resonant mode
interaction (Moskalik 1985) is consistent with observations of this
phenomenon in $\delta$ Scuti stars (e.g. Handler et al. 1998, 2000),
whereas frequency changes of rapidly oscillating Ap stars may be
attributed to variations in the magnetic field (Kurtz et al. 1994, 1997).

Time-resolved photometric observations of pulsating (pre-)white dwarf
stars revealed that they are no exception in this respect. This bears a
potentially enormous astrophysical reward: although these stars may only
make part of their pulsation spectra observable to us at a given time,
they may reveal their complete mode spectrum when observed persistently.
Kleinman et al. (1998), Bond et al. (1996) and Vauclair et al. (2002) took
advantage of this possibility and could then make seismic analyses for a
pulsating DA white dwarf (G29-38) and two pulsating central stars of
planetary nebulae (NGC 1501, RXJ 2117+3412). Without their amplitude
variability, these stars would still be poorly understood.

Published reports of amplitude and frequency variations are still sparse
for the Helium-atmosphere pulsating DB white dwarf stars (DBVs, see
Bradley 1995 for a review), but so are their time-series photometric
observations, mainly due to their faintness (most DBVs are around 16$^{\rm
th}$ magnitude). The glaring ($B=13.5$) exception is the prototype DBV GD
358 for which a plethora of measurements, including three Whole Earth
Telescope (WET, Nather et al. 1990) runs, is available. Although the mode
amplitudes of GD 358 vary, the associated pulsation frequencies seem
reasonably stable (Kepler et al. 2002). Very recently, some amplitude and
frequency variations have also been reported for the DBVs CBS 114 and PG
1456+103 (Handler, Metcalfe \& Wood 2002).

We have started a systematic observing program to obtain reliable
frequency analyses of the mode spectra of all pulsating DB white dwarfs.
Our measurements consist of extensive single-site observations, which can
suffice for very simply-behaved objects (Handler 2001), or low-priority
WET observations (this paper), or even full worldwide multisite campaigns.

\vspace{4mm}

\section{Observations and reductions}

The pulsating DB white dwarf stars KUV\,05134+2605 and PG\,1654+160 were
chosen as secondary target stars for the WET runs Xcov20 and Xcov21 during
November 2000 and April 2001, respectively. Such secondary programme stars
are observed by the network if the primary target is not observable or if
two telescopes are on line and the larger one already measures the primary
or if the observing method at a certain site is not suitable for the
primary (e.g. CCDs are not proper instruments for very bright stars).
Although the temporal coverage of the variations of a secondary target is
usually considerably poorer than that of the primary, the resulting data
sets are often quite valuable (see Handler et al. 1997 for an example).

In addition to the WET measurements, we acquired single-site observations
of KUV\,05134+2605 and PG\,1654+160 before and/or after the main data
stream. In an effort to understand these two stars to the limits currently
possible, we also (re)analysed all available published and unpublished
measurements. The time-series photometric data at our disposal are listed
in Tables 1 and 2.

\begin{table}
\caption[]{Time-series photometry of KUV\,05134+2605. The first part of
the table contains the WET measurements, the second part lists additional
single-site observations, and the third part contains the available
discovery data (Grauer et al. 1989). Runs marked with asterisks were
obtained with a CCD.}
\begin{center}
\scriptsize
\begin{tabular}{llccc}
\hline
Run Name & Obs./Tel. & Date & Start & Length \\
 & & (UT) & (UT) & (h) \\
\hline
asm-0080 & McD 2.1m & 2000 Nov 20 & 11:32:40 & 0.90\\
asm-0082 & McD 2.1m & 2000 Nov 21 & 10:35:30 & 1.78\\
joy-003 & McD 2.1m & 2000 Nov 23 & 10:20:00 & 2.16\\
joy-013 & McD 2.1m & 2000 Nov 26 & 08:52:16 & 3.55\\
joy-017 & McD 2.1m & 2000 Nov 27 & 10:08:40 & 2.38\\
joy-021 & McD 2.1m & 2000 Nov 28 & 07:16:20 & 5.10\\
joy-026 & McD 2.1m & 2000 Nov 29 & 09:48:50 & 2.77\\
sara-0054$^{\ast}$ & SARA 0.9m & 2000 Nov 30 & 04:53:00 & 8.49\\
joy-029 & McD 2.1m & 2000 Nov 30 & 09:51:20 & 2.50\\
joy-032 & McD 2.1m & 2000 Dec 01 & 09:10:00 & 3.44\\
sara-0055$^{\ast}$ & SARA 0.9m & 2000 Dec 03 & 05:18:20 & 8.01\\
sara-0057$^{\ast}$ & SARA 0.9m & 2000 Dec 05 & 03:21:00 & 5.47\\
tsm-0088 & McD 2.1m & 2000 Dec 05 & 09:46:00 & 2.67\\
\hline
KU1229OH & OHP 1.9m & 1992 Dec 29 & 18:23:00 & 5.00\\
KU1230OH & OHP 1.9m & 1992 Dec 30 & 17:50:00 & 4.50\\
gh-0484$^{\ast}$ & SAAO 1.0m & 2000 Oct 04 & 00:45:26 & 2.55\\
gh-0486$^{\ast}$ & SAAO 1.0m & 2000 Oct 05 & 00:42:07 & 2.02\\
gh-0488$^{\ast}$ & SAAO 1.0m & 2000 Oct 06 & 00:36:43 & 2.76\\
gh-0492$^{\ast}$ & SAAO 1.0m & 2000 Oct 09 & 00:15:49 & 3.03\\
gh-0493$^{\ast}$ & SAAO 0.75m & 2001 Jan 30 & 19:01:40 & 2.40\\
gh-0496$^{\ast}$ & SAAO 0.75m & 2001 Jan 31 & 18:57:10 & 2.40\\
gh-0499$^{\ast}$ & SAAO 0.75m & 2001 Feb 01 & 18:40:15 & 2.62\\
\hline
adg-0076 & MtB 1.5m & 1988 Oct 12 & 09:37:20 & 2.43\\
adg-0077 & MtB 1.5m & 1988 Oct 13 & 08:23:10 & 1.67\\
adg-0080 & MtB 1.5m & 1988 Oct 14 & 06:52:30 & 1.10\\
adg-0081 & MtB 1.5m & 1988 Oct 14 & 10:19:20 & 1.86\\
adg-0082 & MtB 1.5m & 1988 Oct 16 & 09:45:10 & 2.48\\
adg-0084 & MtB 1.5m & 1988 Oct 18 & 07:26:00 & 1.98\\
\hline
Total WET & & & & 49.22\\
Grand total & & & & 88.02\\
\hline
\end{tabular}
\normalsize
\end{center}
Observatory codes: McD = McDonald Observatory (USA), SARA = Southeastern
Association for Research in Astronomy Observatory (USA), OHP = Observatoire
de Haute-Provence (France), SAAO = South African Astronomical Observatory,
MtB = Steward Observatory (Mt. Bigelow site, USA)
\end{table}

\begin{table}
\caption[]{Time-series photometry of PG\,1654+160. The first part of the
table contains the WET measurements, the second part lists additional
single-site observations, and the third part reports the available
discovery data (Winget et al. 1984). Runs marked with asterisks were
obtained with a CCD.}
\begin{center}
\scriptsize
\begin{tabular}{llccc}
\hline
Run Name & Obs./Tel. & Date & Start & Length \\
 & & (UT) & (UT) & (h) \\
\hline
sara-0081$^{\ast}$ & SARA 0.9m & 2001 Apr 17 & 10:53:30 & 1.36\\
luc02a & LNA 1.6m & 2001 Apr 20 & 04:36:10 & 0.36\\
luc03a & LNA 1.6m & 2001 Apr 20 & 05:51:10 & 1.13\\
luap21c & LNA 1.6m & 2001 Apr 21 & 03:12:00 & 2.48\\
luap21d & LNA 1.6m & 2001 Apr 21 & 05:43:40 & 2.32\\
sara-0083$^{\ast}$ & SARA 0.9m & 2001 Apr 21 & 06:52:25 & 5.33\\
mdr160 & CTIO 1.5m & 2001 Apr 21 & 08:38:50 & 1.56\\
luap22c & LNA 1.6m & 2001 Apr 22 & 07:10:40 & 1.11\\
mdr163 & CTIO 1.5m & 2001 Apr 22 & 09:05:20 & 1.05\\
luap23b & LNA 1.6m & 2001 Apr 23 & 06:42:00 & 1.75\\
sara-0085$^{\ast}$ & SARA 0.9m & 2001 Apr 23 & 09:48:50 & 2.26\\
luap24c & LNA 1.6m & 2001 Apr 24 & 02:58:00 & 1.35\\
luap24d & LNA 1.6m & 2001 Apr 24 & 04:29:00 & 0.89\\
mdr166 & CTIO 1.5m & 2001 Apr 24 & 07:50:30 & 2.33\\
luap25b & LNA 1.6m & 2001 Apr 25 & 03:27:00 & 2.13\\
mdr169 & CTIO 1.5m & 2001 Apr 25 & 06:37:00 & 3.45\\
gh-0508 & SAAO 1.9m & 2001 Apr 26 & 02:06:40 & 1.94\\
luap26b & LNA 1.6m & 2001 Apr 26 & 03:01:00 & 5.40\\
NOTkd25b$^{\ast}$ & NOT 2.6m & 2001 Apr 26 & 03:29:00 & 2.46\\
mdr172 & CTIO 1.5m & 2001 Apr 26 & 09:03:40 & 1.14\\
gh-0509 & SAAO 1.9m & 2001 Apr 27 & 00:59:40 & 2.83\\
NOTkd26c$^{\ast}$ & NOT 2.6m & 2001 Apr 27 & 01:17:20 & 4.49\\
luap27b & LNA 1.6m & 2001 Apr 27 & 03:12:00 & 3.49\\
luap27c & LNA 1.6m & 2001 Apr 27 & 06:52:00 & 1.46\\
mdr175 & CTIO 1.5m & 2001 Apr 27 & 09:05:20 & 1.06\\
sara-0092$^{\ast}$ & SARA 0.9m & 2001 Apr 28 & 06:22:50 & 5.73\\
mdr178 & CTIO 1.5m & 2001 Apr 28 & 07:51:40 & 2.30\\
sara-0095$^{\ast}$ & SARA 0.9m & 2001 Apr 29 & 06:05:25 & 5.96\\
gh-0517 & SAAO 1.9m & 2001 May 01 & 02:07:30 & 1.90\\
NOTkd30c$^{\ast}$ & NOT 2.6m & 2001 May 01 & 03:49:40 & 1.84\\
NOTke02a$^{\ast}$ & NOT 2.6m & 2001 May 02 & 23:08:50 & 5.11\\
NOTke03d$^{\ast}$ & NOT 2.6m & 2001 May 04 & 02:22:42 & 3.33\\
\hline
r3076 & McD 2.1m & 1985 Jun 21 & 03:39:00 & 6.61\\
sjk-0317 & McD 0.8m & 1994 Apr 06 & 06:40:30 & 5.14\\
sjk-0319 & McD 0.8m & 1994 Apr 07 & 06:28:00 & 5.42\\
pi-001$^{\ast}$ & PO 1.0m & 2001 May 14 & 21:27:53 & 4.59\\
pi-002$^{\ast}$ & PO 1.0m & 2001 May 16 & 22:21:34 & 3.61\\
pi-003$^{\ast}$ & PO 1.0m & 2001 Jun 25 & 19:49:39 & 1.89\\
pi-004$^{\ast}$ & PO 1.0m & 2001 Jun 26 & 20:47:06 & 4.01\\
pi-005$^{\ast}$ & PO 1.0m & 2001 Jun 27 & 20:51:02 & 2.50 \\
\hline
r2817 & McD 2.1m & 1983 Aug 02 & 05:32:46 & 1.43\\
r2819 & McD 2.1m & 1983 Aug 04 & 03:27:13 & 0.47\\
r2822 & McD 2.1m & 1983 Aug 07 & 03:16:00 & 3.49\\
\hline
Total WET & & & & 81.30 \\
Grand total & & & & 120.46\\
\hline
\end{tabular}
\normalsize
\end{center}
Observatories: SARA = Southeastern Association for Research in
Astronomy Observatory (USA), LNA = Osservat\'orio do Pico dos Dias
(Brazil), CTIO = Cerro Tololo Interamerican Observatory (Chile), SAAO =
South African Astronomical Observatory, NOT = Nordic Optical Telescope
(Tenerife), McD = McDonald Observatory (USA), PO = Piszkestet\"o 
Observatory (Hungary)
\end{table}

Most of the observations consisted of multichannel high-speed
photoelectric photometry with 10-s integrations (see Kleinman, Nather \&
Philips 1996 for more information). Channel 1 measured the programme star,
channel 2 measured a local comparison star, and channel 3 simultaneously
recorded sky background. If no third channel was available, the
measurements were irregularly interrupted to measure sky. Data reduction
was performed with a standard procedure, as e.g. described by Handler et
al. (1997).

Our CCD measurements were acquired with a number of different photometers
-- which we do not describe in detail here. The observations were
optimised to acquire at least two local comparison stars in the same field
as the target by minimising the readout time, ensuring a duty cycle as
high as possible. In this way, consecutive data points were obtained in 10
- 30\,s intervals, depending on the instrument.

CCD data reduction comprised correction for bias, dark counts and
flat field. Photometric measurements on these reduced frames were made with
the programs MOMF (Kjeldsen \& Frandsen 1992) or RTP ({\O}stensen 2000),
and differential light curves were created. No variability of any star
other than the targets in the different CCD fields was found, and the
comparison star ensemble resulting in the lowest scatter in the target
star light curves was chosen.

At this point it should be noted that PG\,1654+160 has a companion star
(Zuckerman \& Becklin 1992) at a separation of about $4\arcsec$ distance
that may affect our measurements. Fortunately for us, this companion is
very red. Consequently, we used red-cutoff filters, e.g. a Schott BG 39
glass, which suppressed the companion's contribution sufficiently (it then
was $\sim$2 mag fainter than the target), but did not waste too many
photons of the target star. This also means that the companion's flux did
not affect the amplitudes of the photoelectrically measured target star
light curves significantly, as all our photomultipliers are
blue-sensitive.

Finally, the times of measurement were transformed to Barycentric Julian
Ephemeris Date (BJED); the barycentric correction was applied point by
point. Finally, some overlapping portions of the combined light curves
were merged, and the reduced time series were subjected to frequency
analyses.

\section{Frequency analysis}

Our frequency analyses were mainly performed with the program {\tt PERIOD
98} (Sperl 1998). This package applies single-frequency power spectrum
analysis and simultaneous multi-frequency sine-wave fitting. It also
includes advanced options, such as the calculation of optimal light-curve
fits for multi-periodic signals including harmonic, combination, and
equally spaced frequencies, which are often found in the analysis of the
light curves of pulsating white dwarf stars.

In one case to be indicated later, this method was supplemented by a
residualgram analysis (Martinez \& Koen 1994), which is based on a
least-squares fit of a sine wave with $M$ harmonics. One advantage of this
method is that alias ambiguities can be evaluated more reliably by the
simultaneous inclusion of the information in the Fourier harmonics.

\subsection{KUV\,05134+2605}

We first analysed the WET measurements of KUV\,05134+2605 with {\tt PERIOD
98}. We computed the spectral window of the data (calculated as the
Fourier Transform of a noise-free sinusoid with a frequency of 1.902 mHz
and an amplitude of 9.7 milli-modulation amplitudes\footnote{One
milli-modulation amplitude is the Fourier amplitude of a signal with a
fractional intensity variation of 0.1\%; it is a standard unit for WET
data analysis.} (mma)) followed by the amplitude spectrum itself. The
results are shown in the upper two panels of Fig.\,1. As the WET
measurements were only acquired from North American observatories (a
result from the star having second priority), the window function is poor.

\begin{figure}
\includegraphics[width=84mm,viewport=-20 00 275 505]{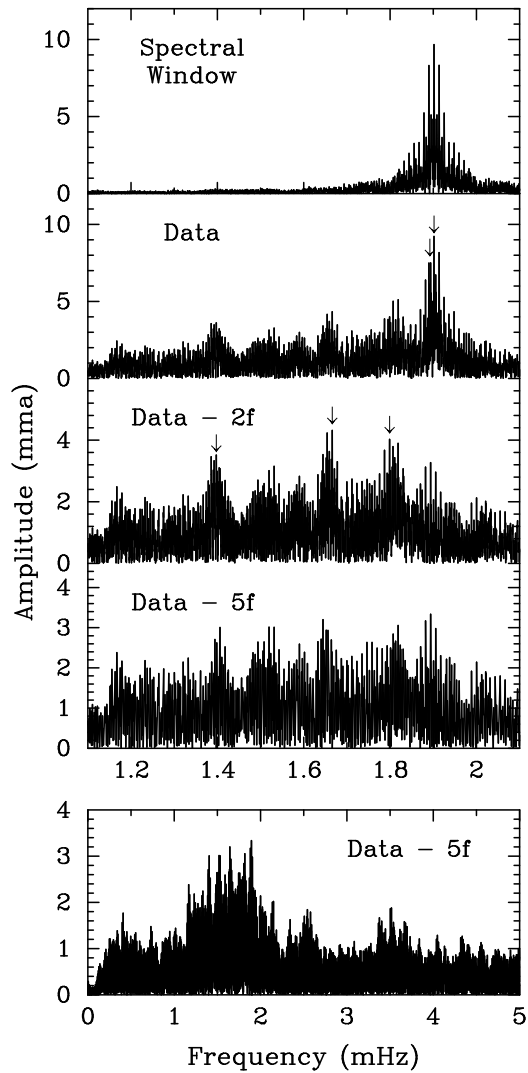}
\caption[]{Spectral window and amplitude spectra of the WET measurements 
of KUV\,05134+2605. Some trial prewhitening of the signals indicated with 
arrows to demonstrate the richness of the frequency spectrum is shown in 
consecutive panels. Due to the poor spectral window, no frequency can be 
determined with certainty, but many signals are definitely present in the 
light curves, as best seen in the lowest panel which shows an extended 
frequency range.}
\end{figure}

Still, we show some prewhitening steps in consecutive panels in Fig.\,1.  
This has been done to indicate the main regions in which pulsational
signals are present, but definite periods cannot be determined. In any 
case, it is suggested that KUV\,05134+2605 has a rich pulsation spectrum.

This is not the only interesting feature of the pulsations of the star: in
the discovery paper (Grauer et al. 1989) it was found to be of much higher
amplitude and longer period compared to its pulsational state during the
WET run. The star has changed from showing dominant pulsations with time
scales of 710\,s and peak-to-peak amplitudes up to 0.2 mag to less than
0.1 mag and a dominant 530-s time scale (Fig.\,2).

\begin{figure}
\includegraphics[width=84mm,viewport=-10 00 275 210]{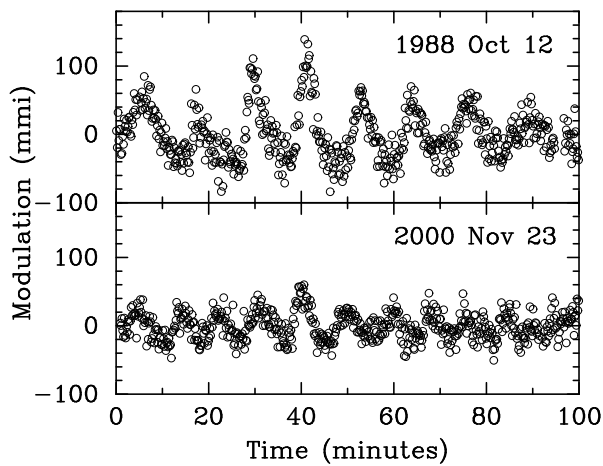}
\caption[]{Upper panel: the discovery light curve of KUV\,05134+2605. 
Lower panel: one of the light curves acquired during the WET run on the 
star. Note the change in the pulsational time scales and amplitudes.}
\end{figure}

\begin{figure}
\includegraphics[width=84mm,viewport=-20 00 275 505]{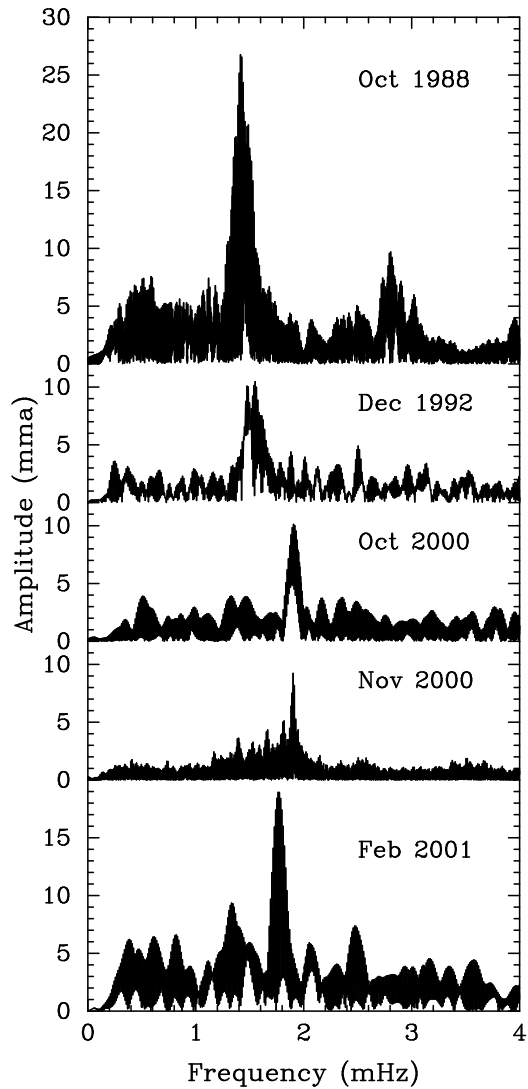}
\caption[]{Amplitude spectra of all available measurements of KUV
05134+2605. The frequencies and amplitudes of the dominant signals are
different in almost every data set.}
\end{figure}

We have therefore calculated amplitude spectra of all the available data
(Fig.\,3). Interestingly, the amplitude spectrum of the star was different
every time it was observed. Besides the data discussed before, the
measurements from 1992 show a dominant variability time scale of around
650\,s, and the February 2001 data have a prevailing time scale of 570\,s.
Only the light curves from October 2000 appear similar to those acquired
by the WET some 6 weeks later, but this actually only applies to the
signal of highest amplitude. We conclude that KUV\,05134+2605 shows
notable amplitude variability on time scales as short as 6 weeks. We
cannot make any statement about frequency/phase variability as our data
sets are too small to distinguish this hypothesis from the effects of
beating between several signals.

We can attempt to construct the complete mode spectrum of the star by
combining the results of the different observing seasons. Of course, all
the data sets are affected by aliasing and no definite periods can be
determined, but approximate periods in the different regions of power in
the Fourier spectra that are separated by more than the width of the
envelope of the corresponding spectral window, can be estimated. We
summarize these results, omitting possible combination frequencies, in
Table 3. The amplitudes in this table must be taken with some caution, as
they could be affected by insufficient frequency resolution in some of the
data sets.

\begin{table}
\caption[]{Dominant signals in the light curves of KUV\,05134+2605 in our
data. The error estimates in the periods include alias ambiguities, and 
amplitudes are listed for completeness.}
\begin{center}
\begin{tabular}{lcc}
\hline
Month/Year & Period (s) & Amplitude (mma)\\
\hline
Oct 1988 & 707 $\pm$ 6 & 25\\
 & 665 $\pm$ 7 & 16\\
 & 777 $\pm$ 8 & 10\\
Dec 1992 & 645 $\pm$ 9 & 10\\
 & 678 $\pm$ 11 & 9\\
Oct 2000 & 525 $\pm$ 7 & 9\\
Nov 2000 & 526 $\pm$ 3 & 9\\
 & 556  $\pm$ 4 & 4\\
 & 600  $\pm$ 4 & 4\\
 & 716  $\pm$ 5 & 3\\
Feb 2001 & 567 $\pm$ 8 & 18\\
 & 757 $\pm$ 14 & 8\\
\hline
\end{tabular}
\end{center}
\end{table}

As already noted, there is no correspondence between the period of the
dominant modes in each of the subsets of data except for the two closest
in time (October/November 2000). It almost appears that we looked at a
different star every time KUV\,05134+2605 was observed!

\begin{figure*}
\includegraphics[width=177mm,viewport=-08 05 500 110]{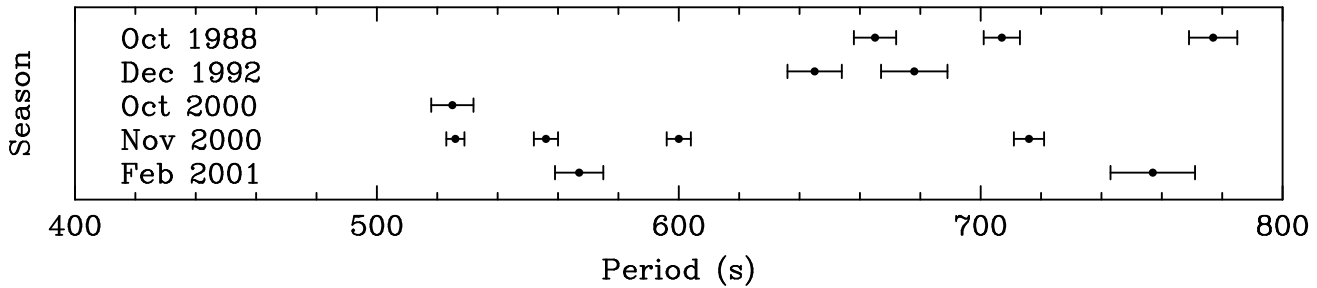}
\caption[]{The variability periods of KUV\,05134+2605 as listed in Table 3.}
\end{figure*}

In any case, we tried to find the signatures of non-radial gravity (g)-mode
pulsations from Table 3, also with the help of Fig.\,4, where we plot the
detected periods over the different observing seasons. However, equally
spaced periods suggestive of the presence of a number of radial overtones
of the same $\ell$ or equally spaced frequencies that might be due to
rotational $m$-mode splitting were not detected.

We can therefore summarize the frequency analysis of KUV\,05134+2605 as
follows: it has a very rich mode spectrum and its pulsation amplitudes are
highly variable. We cannot find a stable pulsational signal which would 
allow us to estimate an evolutionary period change. Only a dedicated 
multisite campaign would help to understand this star.

\subsection{PG\,1654+160}

We again start the frequency analysis with the WET measurements. The 
spectral window and amplitude spectrum of these data are shown in Fig.\,5. 
Although the amplitude spectrum does not appear very complicated, attempts 
to determine the underlying variations by prewhitening result in a large 
number of signals that seem to be present.

\begin{figure}
\includegraphics[width=84mm,viewport=-20 00 275 410]{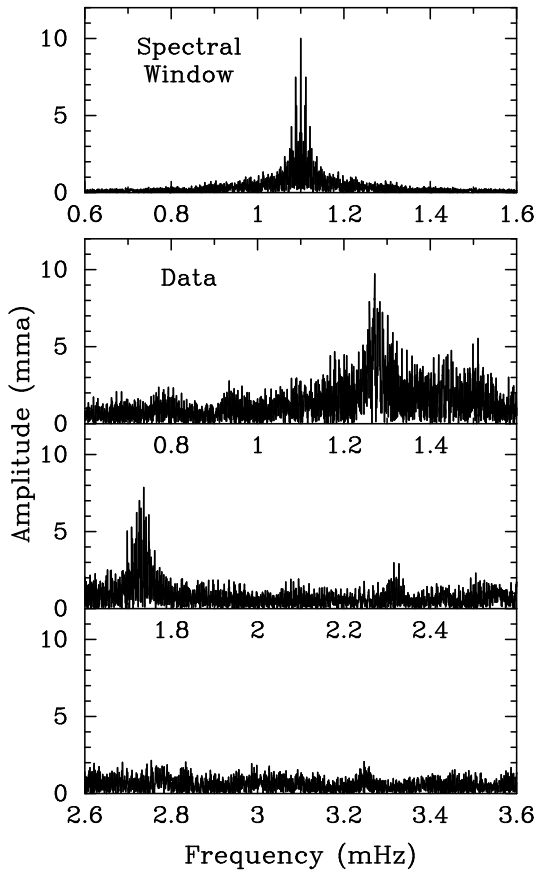}
\caption[]{Spectral window and amplitude spectra of the WET measurements 
of PG\,1654+160.}
\end{figure}

However, assuming that we deal with normal-mode pulsations of the star,
the number of signals becomes unrealistically large, and some of the
frequencies found that way are too closely spaced to be resolved within
our data set. All this suggests that the amplitude spectrum of
PG\,1654+160 was not stable throughout the observations.

Consequently, we attempted to follow the suspected amplitude and frequency
variability throughout the data set with various methods but again had to
realize that our temporal coverage is insufficient for a detailed
analysis. Some results can however be obtained:

\begin{itemize}
\item The longer period pulsations ($P \simgt 700$ s) show a larger degree 
of instability;
\item The amplitude spectrum was more stable during the second part of the 
run (beginning with April 25);
\item Amplitude variability alone is insufficient to account for the 
observed variations; the pulsation frequencies also appear somewhat 
variable;
\item The periods of the strongest signals can be determined, albeit with 
large errors due to the instability and aliasing.
\end{itemize}

The periods we could determine are listed in Table 4, together with the 
results from the other data sets to which we now turn. 

\begin{table}
\caption[]{Dominant signals in our light curves of PG\,1654+160. The error
estimates in the periods include alias ambiguities, and amplitudes are 
listed for completeness.}
\begin{center}
\begin{tabular}{lcc}
\hline
Month/Year & Period (s) & Amplitude (mma)\\
\hline
Aug 1983 & 577 $\pm$ 6 & 24 \\
 & 842 $\pm$ 13 & 15\\
June 1985 & 777: & 14 \\
 & 756: & 11\\
 & 705 $\pm$ 6 & 10 \\
 & 878 $\pm$ 8 & 10 \\
 & 817 $\pm$ 7 & 8 \\
 & 656 $\pm$ 5 & 6 \\
Apr 1994 & 927 $\pm$ 10 & 31 \\
 & 854 $\pm$ 16 & 12 \\
Apr/May 2001 (WET) & 781 $\pm$ 6 & 10 \\
 & 579 $\pm$ 3 & 8 \\
 & 662.0 $\pm$ 0.5 & 5 \\
 & 700 $\pm$ 1 & 4 \\
 & 833 $\pm$ 4 & 4 \\
 & 431 $\pm$ 1 & 3 \\
May 2001 & 791 $\pm$ 15 & 21 \\
 & 575 $\pm$ 6 & 20 \\
 & 913 $\pm$ 15 & 19 \\
June 2001 & 697 $\pm$ 5 & 19 \\
 & 658 $\pm$ 5 & 12 \\
\hline
\end{tabular}
\end{center}
\end{table}

\begin{figure}
\includegraphics[width=83mm,viewport=-20 05 275 542]{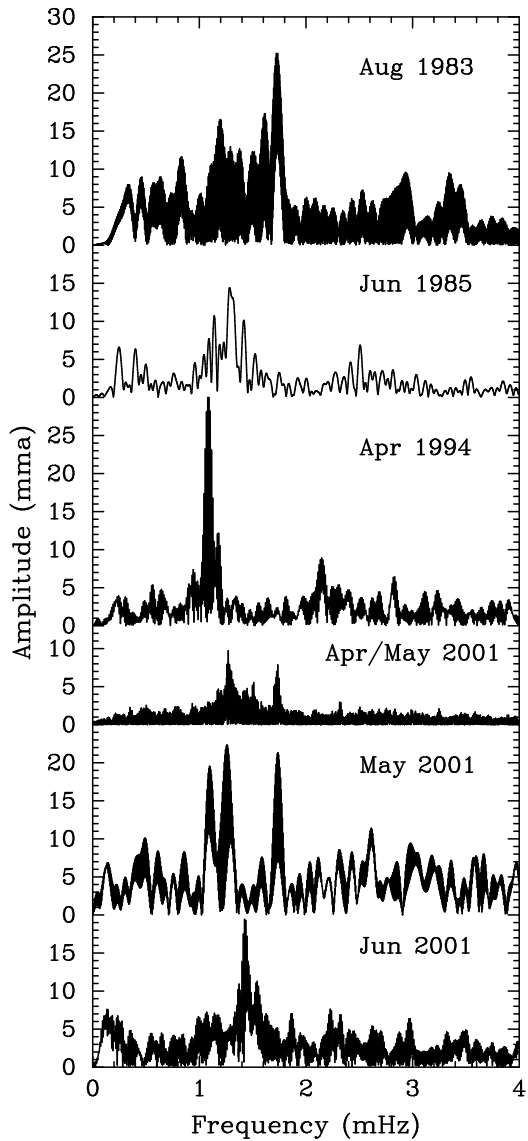}
\caption[]{Amplitude spectra of all available measurements of PG\,1654+160. 
This star also shows conspicuous amplitude variability.}
\end{figure}

In the same fashion as in the previous section, we computed amplitude
spectra of all our data sets over the years. We show them in Fig.\,6 which
demonstrates that the pulsational behaviour of PG\,1654+160 is also highly
variable in time; a comparison of light curves is shown in Fig.\,7. The
time scales of the amplitude variations of PG\,1654+160 can be as short as
two weeks: the strong 913-s signal in the May 2001 data was absent in the
previous WET data.

\begin{figure}
\includegraphics[width=84mm,viewport=-10 00 275 210]{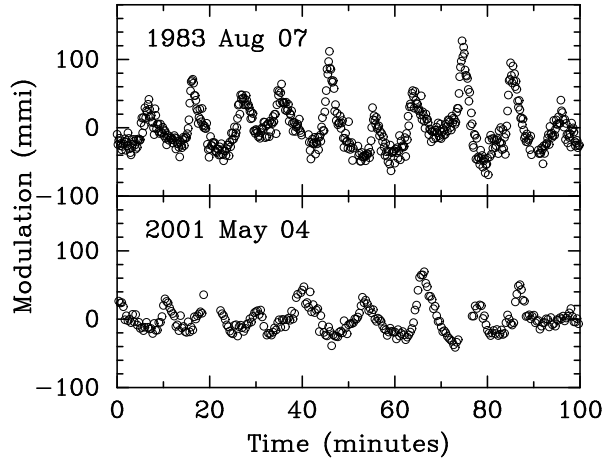}
\caption[]{Upper panel: the discovery light curve of PG\,1654+160.
Lower panel: one of the light curves acquired during the WET run on the
star. Again, the pulsational time scales and amplitudes changed 
considerably.}
\end{figure}

We determined the dominant periods in the different data sets, and
summarize them in Table 4. Again, the amplitudes may be affected by
resolution problems, and possible combination frequencies were excluded. 
In addition, there is good evidence for more signals being present in 
several of the data sets, but it is not possible to determine their 
periods and amplitudes reliably.

Some comments are necessary: the two strongest modes in the single-night
data set from June 1985 are not resolved, which is why we cannot determine
error estimates for their periods and thus disregard them for the
following analysis. The errors on the other frequencies in this data set
were assumed to be $1/4T$, where $T$ is the length of the run. For
the other data sets, the errors on the periods include some possible alias
ambiguities. In the April 1994 data, the 2-f harmonic of the dominant
periodicity is also present. We therefore used the residualgram method (as
described before) with $M=2$ to obtain a more reliable determination of
this period before searching for more signals.

It is interesting to note that some periodicities in Table 4 occur in more
than one data set. We have displayed these results graphically in Fig.\,8,
where we again show the detected periods over the different observing
seasons. We note that the shorter periods ($P<800$\,s) in the light curves
of PG\,1654+160 appear very consistently in the same regions, whereas the
longer periods do not show that much regularity. However, the errors in
the determination of those periods are also larger. Finally, the
previously mentioned shorter-period signals seem to have an approximately
equidistant spacing of about 40 s, and again our data show no sufficiently
stable modes to estimate the evolutionary period change rate.

\begin{figure*}
\includegraphics[width=177mm,viewport=-05 05 507 130]{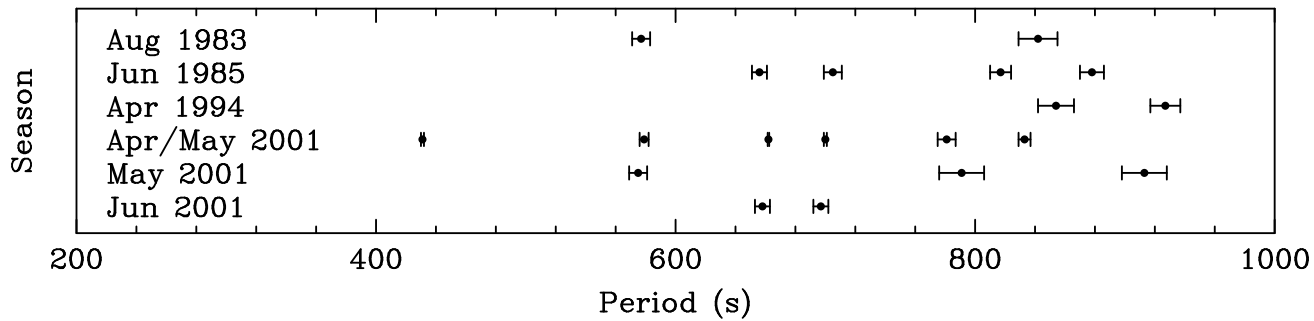}
\caption[]{The variability periods of PG\,1654+160 as listed in Table 4.
Signals with periods below 800~s occur at the same frequencies whenever 
detected, and they are spaced by integer multiples of about 40~s.}
\end{figure*}

\section{Discussion}

We have shown that the amplitude spectra of both KUV\,05134+2605 and 
PG\,1654+160 are variable in time. Whereas we cannot find an underlying
pattern in the periods of KUV\,05134+2605, the roughly equidistant
spacings within the shorter periods of PG\,1654+160 suggests the presence
of a number of radial overtones of g-modes. The size of this average
period separation ($\approx 40$\,s) is consistent with the expected mean
period spacing of a normal-mass ($\approx 0.6 M_{\sun}$) DBV white dwarf
pulsating in $\ell=1$ modes (see, e.g. Bradley, Winget \& Wood 1993).

A comparison of the individual mode periods of the known $\ell=1$ pulsator
GD 358 (Winget et al. 1994, Vuille et al. 2000) and those of another DBV,
CBS 114 (Handler et al. 2002), with that of PG\,1654+160 also supports
this interpretation. However, the number of available observed modes of
PG\,1654+160 is insufficient for seismic model calculations,
and the uncertainties of their periods are too large.

What may be the cause of the amplitude (and possibly also frequency)  
variability in the two stars? As neither has been reported to be magnetic
in the literature, interaction between the different pulsation modes
remains the most promising hypothesis for an explanation.

In this case, the time scale of the amplitude variability is expected to
be of the order of the inverse growth rates of the affected modes. Growth
rates are not very well known for pulsating white dwarfs, but it is clear
that longer-period modes have larger growth rates than shorter-period
ones. Detailed growth-rate calculations (Dolez \& Vauclair 1981) imply
that amplitude variability may occur on time scales down to about one
week.

These theoretical predictions are consistent with our observations, at
least as far as we can tell. The longer-period modes of PG\,1654+160
indeed seem to vary more rapidly in amplitude than the ones at shorter
period (it is interesting to note that Kepler et al.\,(2002) made the same
observation for GD\,358), as implied by our attempts to trace these
variations. The time scale of the amplitude variability of both stars
also appears to be of the expected order of magnitude. However, we must
admit that the temporal distribution of our data is such that we can only
detect variations on just these time scales. Hence, the agreement we find
can at best be regarded as qualitative.

\section{Summary and conclusions}

We have carried out new Whole Earth Telescope measurements of the two
pulsating DB white dwarf stars KUV\,05134+2605 and PG\,1654+160 which were
supplemented by single-site data. We also re-analysed all available
archival measurements of the two stars. 

We showed, for the first time, that both have rich pulsational mode
spectra, and that the pulsation amplitudes of both stars are highly
variable in time; PG\,1654+160 may show some frequency variability in
addition. Beating of multiple pulsation modes cannot explain all our
observations, as the observed amplitude and frequency variability is too
complex for such an interpretation; hence it must be intrinsic. The
pronounced amplitude variations made it impossible to find stable modes to
determine the evolutionary period change rates of the two stars.

Whereas there seems no systematic pattern in the periods of
KUV\,05134+2605 we measured, some of the modes of PG\,1654+160 appear at
the same periods whenever detected. The spacing of these periods, around
40\,s, suggests that they are probably caused by non-radial gravity-mode
pulsations of spherical degree $\ell=1$ in a normal-mass DBV white dwarf.

The amplitude variabilities of both stars could be followed by means of
the pre- and post-WET observations that were therefore essential for this
work. Their time scales are consistent with theoretical predictions of
resonant mode coupling. This conclusion is however weakened by the
temporal distribution of our data, which favour the detection of just
those variability time scales.

Before a more detailed investigation of the amplitude variations of these
two stars can be performed (e.g. to guide theoretical work in this
direction, Buchler, Goupil \& Serre 1995), mode identifications and an
improved sampling of the temporal behaviour of the pulsations through
continued single-site measurements are desirable. Given the qualitative
similarity of PG\,1654 and the ``typical'' DBV GD 358, we expect that the
same nonlinear mode coupling and amplitude modulation mechanisms are at
work in both stars. Having very rich mode spectra, both KUV\,05134+2605
and PG 1654+160 are also attractive targets for future extensive multisite
campaigns.

\section*{ACKNOWLEDGEMENTS}

We acknowledge support from Iowa State University, in part through NSF
Grant AST-9876655. M.\,Cunha is supported through the grants PD/18893/98,
of FCT-Portugal, and POCTI/1999/FIS/34549 approved by FCT and POCTI, with
funds from the European Community programme FEDER. P.\,Moskalik
acknowledges partial financial support by the Polish KBN grant
5\,P03D\,012\,20.

\bsp

\end{document}